\documentclass[fleqn,usenatbib]{mnras}

\usepackage{newtxtext,newtxmath}
\usepackage[T1]{fontenc}

\DeclareRobustCommand{\VAN}[3]{#2}
\let\VANthebibliography\thebibliography
\def\thebibliography{\DeclareRobustCommand{\VAN}[3]{##3}\VANthebibliography}

\usepackage{graphicx}	% Including figure files
\usepackage{amsmath}	% Advanced maths commands
\usepackage{siunitx}
\usepackage{booktabs}
\usepackage{lipsum}

\DeclareSIUnit \year {yr}
\DeclareSIUnit \parsec {pc}
\DeclareSIUnit \pixel {pix}
\DeclareSIUnit \Mjup {\ensuremath{M_\mathrm{Jup}}}

\newcommand{\software}[1]{\textsc{#1}}
\title[New eclipse timings for 2M1510\,AB]{A new photometric ephemeris for the 2M1510\,AB double brown dwarf eclipsing binary system}

\author[S.T. Millward \& V. Kunovac]{Seb T. Millward$^{1}$\thanks{Contact e-mail: \href{mailto:Seb.Millward@warwick.ac.uk}{Seb.Millward@warwick.ac.uk}} and 
Vedad Kunovac$^{1,2}$%
\\\\
$^{1}$Department of Physics, University of Warwick, Coventry, CV4 7AL, UK \\
$^{2}$Centre for Exoplanets and Habitability, University of Warwick, Coventry, CV4 7AL, UK \\
}
\date{Accepted XXX. Received YYY; in original form ZZZ}

\pubyear{\the\year{}}

\begin{document}
\label{firstpage}
\pagerange{\pageref{firstpage}--\pageref{lastpage}}
\maketitle

\begin{abstract}
% Eclipsing brown dwarfs are important calibrators of sub-stellar evolution models which are used to infer the characteristics of directly imaged brown dwarfs and giant exoplanets.
% Only two double brown dwarf eclipsing binary systems are known, among them is 2MASS J15104786-2818174, or 2M1510\,AB for short, which was published in 2020 with a poorly constrained orbital period.
% Here we analyse TESS full-frame image (FFI) photometry of this faint ($T_\mathrm{mag} = 15.9$) binary and detect a significant ($\qty{>10}\!\sigma$) periodic signal spanning TESS Cycles 1-7 that is consistent with the previous data. We refine the orbital period to $P = \qty[separate-uncertainty=false]{20.897771 \pm 0.000044}{\day}$ -- reducing its present-day uncertainty from \qty{18}{\hour} to \qty{8}{\minute}. Our work is crucial for scheduling follow-up observations of this system for detailed study with e.g. JWST and other photometric facilities. We also find that the recent orbital solution from Doppler data presented by \citet{baycroft2025} is not consistent with existing photometry. A timing offset in the Doppler data produced a spurious signal mimicking retrograde apsidal precession, from which the claimed circumbinary planet 2M1510\, ABb was inferred. Using the corrected data we were still unable to reconcile the radial velocity data with the photometry, suggesting that the radial velocity uncertainties are significantly underestimated, and that the circumbinary planet 2M1510\,ABb is a false positive.
Eclipsing brown dwarfs are important calibrators of sub-stellar evolution models used to infer the characteristics of directly imaged brown dwarfs and giant exoplanets. Only two double brown dwarf eclipsing binary systems are known, among them 2MASS J15104786-2818174 (2M1510\,AB), published in 2020 with a poorly constrained orbital period. Here we analyse TESS full-frame image (FFI) photometry of this faint ($T_\mathrm{mag}=15.9$) binary and detect a significant ($\qty{>10}\!\sigma$) periodic signal spanning TESS Cycles 1–7, consistent with previous data. We refine the orbital period to $20.897782 \pm 0.000036$ d, reducing its present-day uncertainty from 18 h to 8 min. Our work is crucial for scheduling follow-up observations of this system for detailed study with other photometric facilities. We also find that a recent orbital solution from Doppler data is inconsistent with existing photometry.
% original
% A timing offset in the Doppler data produced a spurious signal mimicking retrograde apsidal precession, from which the claimed circumbinary planet 2M1510 ABb was inferred. Using the corrected data we were unable to reconcile the radial velocity data with the photometry, suggesting that the radial velocity uncertainties are significantly underestimated, and that the circumbinary planet 2M1510 ABb is a false positive.
% updated
A timing offset in the Doppler data may have produced a spurious signal mimicking retrograde apsidal precession, from which the claimed circumbinary planet 2M1510 ABb was inferred. From our best attempt at correcting the data we were unable to reconcile the radial velocity data with the photometry, suggesting that the radial velocity uncertainties are underestimated, and that the circumbinary planet 2M1510 ABb may be a false positive.

% We conclude that the circumbinary planet 2M1510\,b is a false positive, and that the uncertainties on the newly published Doppler data are significantly underestimated -- making it challenging to infer the existence of any circumbinary planet based on existing radial velocity measurements.

% We verify their results but trace the discrepancy to a conversion error between JD and MJD in their data, which created a signal that manifests as a negative apsidal precession rate from which the purported circumbinary planet was inferred. Using the corrected timestamps we were still unable to reconcile the radial velocity data with the photometry.
% While we cannot rule out additional Keplerian signals or secular effects in their (epoch-corrected) data,
% We conclude that the uncertainties on their derived Doppler data are significantly underestimated which may have important consequences for the existence of any purported circumbinary planet based on existing radial velocity measurements.
\end{abstract}


%%%%%%%%%%%%%%%%%%%%%%%%%%%%%%%%%%%%%%%%%%%%%%%%%%

%%%%%%%%%%%%%%%%% BODY OF PAPER %%%%%%%%%%%%%%%%%%

\section{Introduction}

%Currently brown dwarfs are defined to have masses between 10 and 80 $M_\mathrm{Jup}$, and therefore are massive enough for the fusion of deuterium but not hydrogen \citep{Chabrier2014}. 
Brown dwarfs are substellar objects with masses in between giant planets and stars, typically \qtyrange[range-units=single]{13}{80}{\Mjup} such that internal temperatures allow for deuterium burning but not hydrogen fusion \citep{Chabrier2014}.
% The masses, radii and ages of directly imaged brown dwarfs and giant planets are typically inferred from sub-stellar evolution models.
The masses, radii and ages of brown dwarfs and giant planets whose orbital inclination are not known -- such as directly imaged objects or double-line binaries -- are typically inferred from sub-stellar evolution models  \citep[e.g. LP-413-53][]{Hsu2023}.
The calibration of such models rely on the direct measurements of these properties for brown dwarfs. However, the list of systems where such measurements can be made remains small.

For a double line binary system with a known orbital inclination, the masses of the binary components can be directly measured. One such method to find orbital inclination is through astrometry, for example \citet{Franson2022}, \citet{Xuan2024} and \citet{Brandt2021}, who all investigate double line binary systems containing brown dwarfs. Alternatively, for double line binary systems with observable eclipses, the shape of the eclipse in the light curve allows for a measurement of the orbital inclination, as well component radii. Only three such double line ultra-cool (spectral type M8 or later) binary systems have been identified: EPIC 2037103875\,AB \citep{David2019}, 2M0535-05\,AB \citep{stassun2006,stassun2007}, and 2M1510\,AB, the system investigated in this paper. Only the latter two are bona-fide double brown dwarf binaries.

The discovery of 2M1510\,AB was announced by \citet{triaud2020} following the observation of the first eclipse on UT night 2017 July 26 with the SPECULOOS Southern Observatory \citep{Delrez2018(spec)}. The eclipse was soon followed up with high-resolution spectroscopy in the red optical and near-infrared with UVES on the VLT and HIRES on Keck, from which line splitting was observed, confirming the system as a double lined binary. However, no further eclipses were observed, and so the main constraint on the orbital period came from the UVES and HIRES radial velocities. This left an uncertainty on the orbital period of roughly $\qty{8.5}{\minute}$ at the time the first set of data was gathered, increasing on each subsequent eclipse of the system by $\qty{2.5}{\hour}$ per year. At time of writing, the uncertainty is roughly $\qty{18}{\hour}$.

In this Letter, we analyse the TESS \citep{Ricker2015(tess)} full-frame image photometry of 2M1510\,AB, identifying 4 new eclipses which reduce the uncertainty on the orbital period. This work will allow for follow-up observations to study this rare system in more detail.
% Further, we investigate discrepancies between our solution and the recent solution presented by \citet{baycroft2025}, who presented evidence of a circumbinary planet orbiting 2M1510\,AB utilising new, more precise Doppler observations.
The Letter is organised as follows: In Section~\ref{section:data} we summarise the system properties and the extraction of the TESS photometry. In Section~\ref{section:methods} we detail our box-least squares (BLS) search for periodic signals in the TESS data, and the modelling of the light curves using an eclipsing binary model with parameter estimation using MCMC. We outline our main results in Section~\ref{section:results}. In Section \ref{section:discussion} we compare our results to previous findings from \citet{triaud2020}, and the new Doppler solutions found by \citet{baycroft2025}. We also discuss differences between the depth of the eclipses identified in SPECULOOS and TESS. Finally, we conclude in Section~\ref{sec:conclusion}.

\section{Data}
\label{section:data}

\begin{figure*}
	\includegraphics{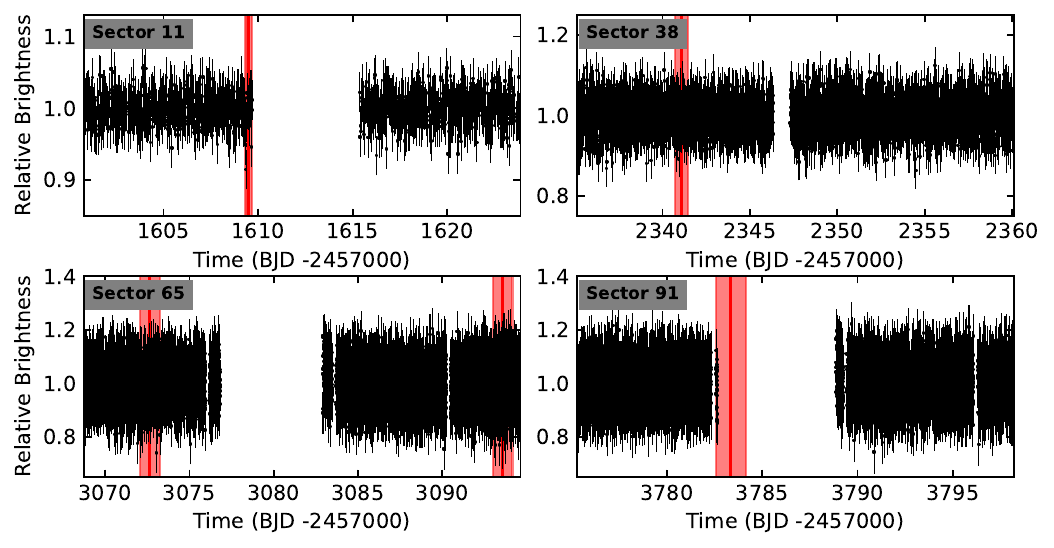}
    \caption{\textbf{Lightcuves of 2M1510AB observed by TESS.} Solid red line indicates position of transits predicted from \citet{triaud2020}. Further red shaded region indicates its $1\sigma$ uncertainty. }
    \label{fig:tess}
\end{figure*}

The 2M1510 system is a hierarchical triple brown dwarf system consisting of an unresolved near equal-mass binary 2M1510\,AB (2MASS J15104786-2818174, Gaia DR3 6212595980928732032), orbited by a spatially resolved third component, 2M1510\,C (2MASS J15104761-2818234, Gaia DR3 6212595980924278144), separated by \qty{6.8}{\arcsec} on the sky or a projected separation of \qty{250}{} astronomical units \citep{gizis2002}. The system is located at a distance of \qty{36.6 \pm 0.3}{\parsec} \citep{collaboration2023} and is a kinematic member of the \qty{45 \pm 5}{\mega\year} Argus moving group \citep{gagne2015}. The inner binary system orbit each other in a \qty{20.9}{\day} eccentric orbit, and a \qty{\sim 4.3}{\percent} deep secondary eclipse was observed with SPECULOOS in 2017. Due to the system being viewed along the line of apsides ($\omega = \qty{-90}{\degree}$) no primary eclipse is visible \citep{triaud2020}.

The 2M1510 system was observed in TESS full-frame images (FFI) in sectors 11, 38, 65 and 91, at exposure times of \qty{30}{\minute} and \qty{10}{\minute} in sectors 11 and 38, and \qty{200}{\second} in sectors 65 and 91. According to the TESS Input Catalog (TIC) the 2M1510AB binary system is designated TIC 61253912 with a $T_\mathrm{mag} = 15.9$, while the tertiary has the TIC identifier 61253915 with a $T_\mathrm{mag} = 17.1$. We use the publicly available software \software{tglc}\footnote{\url{https://github.com/TeHanHunter/TESS_Gaia_Light_Curve?tab=readme-ov-file}} provided by the TESS-Gaia Light Curve project \citealt{han2023} to download the available light curves for the binary system, which can produce light curves down to 16th TESS magnitude. The \software{tglc} software produces light curves by forward modelling the FFIs with the effective point-spread function (ePSF) based on position and magnitude data from Gaia DR3. As a consequence, light curves are automatically corrected for contamination from nearby stars.

Using their FFI forward model, light curves can either be generated from PSF photometry or aperture photometry. For faint stars in sparse fields like ours the PSF method generally yields better precision \citep[][Figure~10]{han2023}. 
To verify this we compute a similar metric to the combined differential photometric precision (CDPP, \citealt{christiansen2012}). In effect, after correcting for low-frequency trends with the biweight filter in \software{wotan} \citep{hippke2019a} and removing outliers with iterative $5\sigma$ clipping, we compute the rolling mean in chunks of 90 minutes; equal to the eclipse duration. We adopt the standard deviation of the rolling means as our CDPP proxy \citep{gilliland2011,vancleve2016}.
Indeed, we find that the PSF light curves led to an improvement between \qtyrange{4}{25}{\percent} in the CDPP relative to the aperture photometry in all sectors except 38, where the CDPP was \qty{6}{\percent} worse. 
\citet{han2023} advocate for a weighted approach using both light curves to get the best precision. Following their work, we calculated the CDPP of the average light curve by increasing the fraction of PSF photometry in steps of 10 percentage points. On average the weighted light curve improved the CDPP by a further \qty{10}{\percent} compared to pure PSF photometry, and we found that the best weights for the PSF photometry were 0.5, 0.4, 0.8 and 0.5 for Sectors 11, 38, 65 and 91, respectively, with the remainder made up from aperture photometry. The CDPP of our final light curves ranged from \qtyrange{1.3}{1.4}{\percent} in a 90 minute window equal to the eclipse duration, which is factor ${>}3$  smaller than the expected eclipse depth. The light curves are shown in Figure~\ref{fig:tess}.

Initially we removed all points flagged by both TESS and \software{tglc} with a quality flag larger than 0. After an initial analysis we noticed that an eclipse occured at the end of Sector 65, indicated by the red vertical region at $\mathrm{BJD} \approx 2460072.2$. This region is flagged as a non-zero quality flag in the \software{tglc} data product, but not in TESS FFI data. To maximize the number of eclipses in our dataset we decided to include this region in our analysis.

\section{Methods}
\label{section:methods}

% bls figure
\begin{figure*}
	\includegraphics{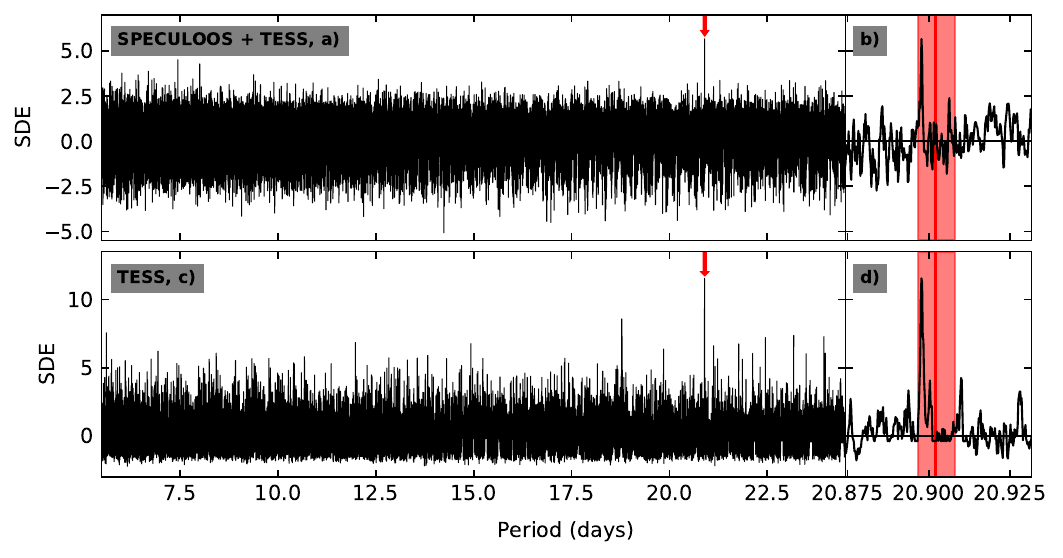}
    \caption{\textbf{Periodograms from Box Least Squares search.} \textbf{a)} is the periodogram from the BLS search using TESS and SPECULOOS. \textbf{b} is magnified for 5 $\sigma$ around the period predicted by \citet{triaud2020}. \textbf{c} and \textbf{d}, same as above, using only TESS data. Red arrows and red line indicate the period predicted by \citet{triaud2020}. Further red shaded region around the red line indicates 1 $\sigma$ uncertainty.}
    \label{fig:bls}
\end{figure*}

To identify approximate eclipse positions within each sector, we extrapolate forwards the eclipse times and errors from the period and secondary transit time of \citet{triaud2020} to each TESS sector. This identifies five transit regions within the TESS data - one in sector 11, one in sector 38, two in sector 65 and one in sector 91 - however, the last transit in sector 91 contains only 32 exposures, covering roughly \qty{4.8}{\percent} of the $1\sigma$ region of interest.
% equal to \qty{1.8}{\hour} compared to the region of interest being \qty{1.55}{\day}.

To search for periodic signals present in the TESS data we implement a box-least squares\footnote{\url{https://docs.astropy.org/en/stable/timeseries/bls.html}} (BLS, \citealt{kovacs2002}) search in the range of 5 to 25 days, searching for eclipses of durations from 60-120 minutes. This search includes the eclipse observed by SPECULOOS as well as all four TESS sectors. To ensure it was not purely a result from the accurate SPECULOOS data being combined with spurious signals in the TESS data, we perform a second BLS search on the TESS data alone. 

We calculate a variation of the signal detection efficiency (SDE) by subtracting a smoothed mean from the raw BLS power spectrum, and then divide the result by the standard deviation. To account for significant noise increase at higher frequencies, we slice the periodogram into 20 different bins and calculate the standard deviation for each bin. The smooth trend of the standard deviation as function of frequency was used as $\sigma_\mathrm{BLS}$.

We perform a least squares fit on the data around the period identified from the BLS search using model light curves generated by the software \software{ellc} \citep{maxted2016}. Following \citet{triaud2020}, we fix the surface brightness ratio $J=0.83$ and radius ratio $k=1$ as they are degenerate for a grazing secondary eclipse. The latter was set due to the mass ratio of the system being indistinguishable from unity, while the former is based on the flux ratio measured in the UVES spectra at \qty{819}{\nano\metre}. The eccentricity $e$ and argument of periastron $\omega$ -- necessary to compute the inferior conjunction times from the time of secondary eclipse $T_\mathrm{sec}$ -- were fixed to the values of \citet{triaud2020} (see Table \ref{tab:variables}). We vary cosine of sky inclination $\cos{i}$, sum of radii ($(R_1 + R_2) / a$), orbital period $P$, and $T_\mathrm{sec}$. We fix the limb darkening parameters, assuming a quadratic limb darkening law, and compute the coefficients for the SPECULOOS $I+z'$ and TESS bands with \software{ldtk}\footnote{\url{https://github.com/hpparvi/ldtk}} \citep{parviainen2015} using stellar parameters from \citet{triaud2020}.

%The surface brightness ratio was fixed at the value from \citet{triaud2020} for SPECULOOS's lightcurve, and varied for the TESS lightcurve's eclipses, such that it could act as a jitter parameter to account for the difference in transit depth between SPECULOOS and TESS.

We then use MCMC sampler \software{emcee} to explore the parameter space \citep{foremanmackey2013}. Walkers are initialised in a normal distribution centred on the values found from the least squares fit.
We use uniform priors, listed in Table~\ref{tab:variables}. We include white noise terms and offset scaling factors for each transit, and account for finite exposure time of the TESS data.
In our fit we include the SPECULOOS eclipse as well as a $1\sigma$ uncertainty region around each TESS eclipse from the \citet{triaud2020} solution, which encompasses several hours around the eclipse times found by the BLS search.
We omit sector 91 as it did not contain any data in the predicted eclipse window from our maximum likelihood solution.
After an initial fit it became clear that the TESS data showed a deeper eclipse than SPECULOOS. Therefore, in the final MCMC, we add a separate brightness ratio for the TESS bandpass to act as a depth parameter, but caution that the value of this parameter is not to be interpreted physically as the discrepancy in the depth is currently unexplained (Section~\ref{section:depthdifference}).

The final run uses 400 walkers. We test it for convergence every 100 iterations, and stop it once the autocorrelation factors had changed by \qty{<1}{\percent}, which results in 54\,600 iterations. We discard the first 52\,600 iterations, and the chains were thinned by a factor of 545 due to autocorrelation, leaving 1200 independent samples for each parameter.

%This process was then repeated for the TESS sectors alone. As both $r_{sum}$ and orbital inclination were now less constrained their prior range was increased so that they only prevented no eclipse from occurring. A Gaussian prior was used for $T_\mathrm{sec}$, using the value and errors from \citet{triaud2020} as mean and standard deviation, so that only solutions with a transit occurring during the SPECULOOS transit were explored. Priors for both fits are shown in \ref{tab:priors}. This run used 400 walkers, and had XXX resultant iterations. The first XXX iterations were discarded, and the chains were thinned by a factor of XXX due to autocorrelation, leaving XXX independent samples for each parameter.

\section{Results}
\label{section:results}

% lightcurve figure
\begin{figure*}
	\includegraphics{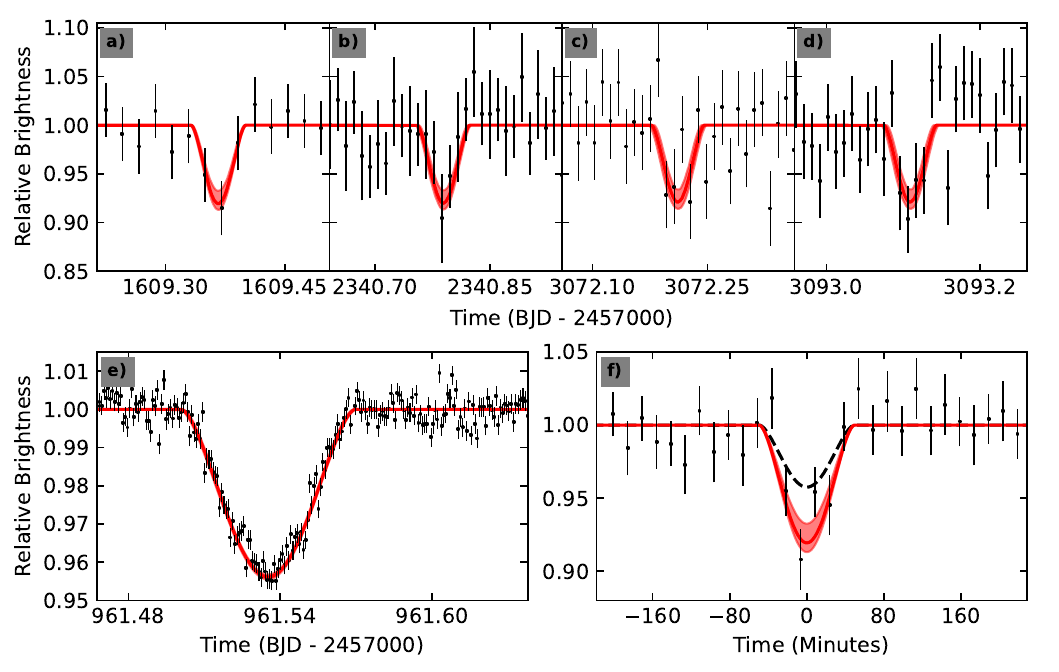}
    \caption{\textbf{Fits for TESS and SPECULOOS data.} \textbf{a)}, \textbf{b)}, \textbf{c)} and \textbf{d)} show the four identified eclipses in the TESS data, binned to 15 minutes. \textbf{e)} shows the eclipse in the SPECULOOS data, and \textbf{f)} shows the phase-folded TESS data of the 4 eclipses, binned to 15 minutes and centred on the mid-eclipse. The red line in each figure indicates the best fit solution from the MCMC run, with a further red shaded region indicating its 1 $\sigma$ uncertainty. The black dotted line in \textbf{f)} represents the best fit solution whilst fixing the surface brightness ratio at 0.827 (i.e. same as SPECULOOS).}
    \label{fig:mcmc}
\end{figure*}

% results table, with Triudad column
\begin{table*} 
	\centering
	\caption{Priors and solutions for the MCMC search, and the solutions found by \citet{triaud2020}.}
	\label{tab:variables}
	\begin{tabular}{lccc}
		\toprule
        \toprule
		Variables and Units &  Priors  & MCMC results & \cite{triaud2020} \\
		\midrule
		Orbital Period ($\mathrm{days}$) &
        $ \mathcal{U}(20.892,20.903) $ & 
        $ 20.897782 \pm 0.000036 $ &
        $ 20.9022_{-0.0056}^{+0.0059} $ \\
        
		$T_\mathrm{sec} (\mathrm{BJD}_\mathrm{TDB} - 2457000)$ & 
        $ \mathcal{U}(961.533,961.537) $ &
        $ 961.53520 \pm 0.00026 $ &
        $ 961.53518_{-0.00061}^{+0.00064} $ \\
        
	    Inclination $(^\circ)$ & 
        $ \mathcal{U}(88,90) $ &
        $ 88.466 \pm 0.030 $ &
        $ 88.5\pm0.1 $ \\
        
        Sum of Radii $(\mathrm{R_\odot})$ &
        $ \mathcal{U}(0.178,0.583) $ &
        $ 0.3145 \pm 0.0057 $ &
        $ 0.3147_{-0.0157}^{+0.0159} $ \\ 

        Surface Brightness Ratio $J_2/J_1 = k^2 f_2/f_1$ &
        Fixed &
        0.827 & 
        $ 0.827 \pm 0.013 $ \\

        Radius Ratio $k=(R_2/R_1)$ &
        Fixed&
        1 &
        1 \\
        
        Eccentricity &
        Fixed&
        0.309& 
        $0.309 \pm 0.022$ \\

        Argument of Periastron $(^\circ$) &
        Fixed&
        -89.9& 
        $-89.9 \pm 3.3$ \\
        
        \bottomrule
		% \hline
	\end{tabular}
\end{table*}

The BLS periodogram from the joint search of SPECULOOS and TESS is shown in the top panel of Figure \ref{fig:bls}. The highest peak in the BLS periodogram corresponds to a period of
$P=\qty{20.89776}{\day}$ with SDE value of 5.7. This period is within $1\sigma$ of the orbital solution based on radial velocity data from \citet{triaud2020}. In the bottom panel of Figure \ref{fig:bls} we show the power spectrum of the BLS search on the TESS data alone, excluding SPECULOOS. This search found six peaks occurring with an SDE value ${>}7$, attributed to the more noisy TESS data. The most significant peak was again identified at $P = \qty{20.89776}{\day}$ with SDE value of 11.5.

Table \ref{tab:variables} presents the 16th, 50th and 84th percentiles of the resultant samples from the MCMC search. The orbital period is $\qty{20.897782 \pm 0.000036}{\day}$, $T_\mathrm{sec}$ is $\qty{2457961.53520 \pm 0.00026}{\day}\,\mathrm{BJD}_\mathrm{TDB}$, orbital inclination is $\qty{88.47 \pm 0.03}{\degree}$ and the sum of radii is $0.3145 \pm 0.0057$. The median model from our posterior is shown in Figure~\ref{fig:mcmc}.
% Although significant noise is present in all four TESS sectors (a, b, c and d), these values produce a good fit to both SPECULOOS' lightcurve (f) and TESS' phase folded lightcurve (e).
We calculate the S/N of the phase folded light curve as as $\mathrm{S/N} = \delta \sqrt{N}/\sigma$ with $\delta$ being the model depth, $N$ is the number of binned exposures inside the eclipse, and $\sigma$ being the individual uncertainty on a single binned \qty{15}{\min} exposure. We find a strong detection with a $\mathrm{S/N} = 9.5$.

The transit depth in the phase folded data is deeper than SPECULOOS, at 8\% compared to 4.2\%, at a significance of $2.6\sigma$. The potential causes of this discussed in Section \ref{section:depthdifference}. To account for this the TESS data was fit with a separate surface brightness ratio value, which acts as a depth rescaling factor for the purposes of our work. It is not to be interpreted as a physical value. This resulted in a surface brightness ratio of $ 6.2_{-3.5}^{+7.3}$ for the TESS bandpass, compared to the fixed 0.83 \citep{triaud2020} for SPECULOOS $I+z'$. Despite its posterior being somewhat broad, the deeper model (red solid line, Figure~\ref{fig:mcmc}) is a better fit to the data than the shallower model produced by using the SPECULOOS value for the surface brightness ratio (black dotted line). Similarly, the fit to the SPECULOOS data which uses the surface brightness ratio from \citet{triaud2020} produces a light curve in a strong agreement.

\section{Discussion}
\label{section:discussion}

\subsection{Comparison to \citet{triaud2020}}
The new ephemeris is in a strong agreement with \citet{triaud2020}, with the orbital period, $T_\mathrm{sec}$, orbital inclination, and the sum of radii all within $1\sigma$ of the solution presented by \citet{triaud2020}. 
Given higher S/N of the SPECULOOS data compared to TESS the shape of the eclipse is strongly constrained by the former,
% A varied surface brightness ratio for the TESS sectors resulted in the shape of the eclipse being strongly constrained by SPECULOOS' lightcurve,
and as a result orbital inclination and the sum of radii are in strong agreement with \citet{triaud2020}.
% but are unable to produce a convincing lightcurve for the TESS data without the use of a jitter parameter.

However, the inclusion of TESS data has reduced the errors on all four of the free parameters - the new $T_\mathrm{sec}$ has an error 2.5 times smaller, the new orbital inclination has an error 3.3 times smaller, and the sum of radii has an error 2.8 times smaller than the errors on the values from \citet{triaud2020}. Finally, the new period we present at $P = \qty{20.897782 \pm 0.000036}{\day}$ has an error of only $\qty{3}{\second}$, reduced from $\qty{8.5}{\minute}$ from \citet{triaud2020}.

\subsection{Comparison to \citet{baycroft2025}}
Recently, \citet{baycroft2025} published 22 new radial velocity (RV) observations of the system using the UVES spectrograph, extending the observed Doppler baseline of the system by up to four years. Using the new data and the 12 previously published epochs from the same instrument, the authors derived significantly more precise RVs by modelling the double-lined spectra using a Gaussian process regression framework \citep{sairam2024}. Compared to the published data, the new analysis reduced the formal uncertainties on the RVs from \qty{\sim 1.5}{\kilo\metre\per\second} to \qty{\sim 50}{\metre\per\second}, an improvement of a factor \qty{\sim 30}{}. \citet{baycroft2025} reported an orbital period of $P = \qty{20.907495 \pm 0.000088}{\day}$, and found that including apsidal precession at a rate of $\dot{\omega} = \qty{-343 \pm 126}{\arcsec\per\year}$ significantly improved the fit. The authors interpreted the retrograde direction of the precession as evidence of a planet orbiting the binary in a polar configuration.

The orbital period we have derived in this work is \qty{14}{\minute} shorter compared to the recent Doppler result from \citet{baycroft2025}, a difference of $110\sigma$. This significant discrepancy prompted us to study the results in more detail. First we compared the binary orbital solution to the published values of \citet{triaud2020}. While the orbital periods agree within errors, $e$ and $\omega$ differ by $2.3\sigma$ and $4.2\sigma$ respectively, which affects the predicted eclipse timings.
We adopted their orbital solution and back-propagated it in time to see if it correctly predicts the secondary eclipse at time $T_\mathrm{sec} = 2457961.53520$ BJD$_\mathrm{TDB}$ that was observed with SPECULOOS in the original discovery paper \citep{triaud2020}. We found that this orbital solution is not consistent with an eclipse at $T_\mathrm{sec}$; the predicted secondary eclipse occurs \qty{13.5}{\hour} after $T_\mathrm{sec}$.

Next we replicated the analysis by using the RV data and epochs (timestamps) which are included in Table 2 and Table 3 in \citet{baycroft2025}, 
% original
% and got identical results to the reported Keplerian fit solution and confirm a negative value of $\dot{\omega}$. 
% updated
and reached $1\sigma$ agreement with most of the reported Keplerian fit solution and confirm a negative value of $\dot{\omega}$. The two most discrepant parameters are the orbital period ($2\sigma$) and $\omega$ ($1.2\sigma$). The differences may arise due to different treatment of nuisance parameters, but we are unable to confirm this based on the available information in the paper. We carried out our analysis with separate RV offsets and white noise terms for each component, and the white noise terms reached a modest \qty{\leq 30}{\metre\per\second} for the two components, suggesting the formal uncertainties are fairly accurately estimated based on the best-fit model.
% the epochs in Table 2 and 3 are rather unusually truncated to three decimal points, and
However, we note that 
22 out of 33 epochs are recorded in MJD despite being labelled BJD -- meaning a 0.5 day offset has not been applied to these affected timestamps. We verified this by downloading the raw UVES data from the ESO archive and checking the timestamps directly.

We find that this temporal shift \textbf{may} have important consequences for the fit. We converted all MJD epochs (in the UTC time scale) from the raw UVES data to BJD$_\mathrm{TDB}$ so that they are consistent, and shifted the epochs to the mid-point of the exposures. We also re-applied the barycentric correction using the corrected epochs, but also verified that whether or not we include this step in our analysis does not affect our final conclusions. We used these new data to fit the same Keplerian model as before, and find no evidence of apsidal precession. This result indicates that the negative value of $\dot{\omega} = \qty{-343 \pm 126}{\arcsec\per\year}$ reported in \citet{baycroft2025} materialised due to inconsistent timestamps, which suggests the purported circumbinary planet 2M1510\,ABb is a false positive. 
However, our model resulted in a poor fit where the uncertainties on the data had to be inflated by \qty{0.6}{\kilo\metre\per\second} in order for the residuals to behave like a normal distribution; an order of magnitude larger than the quoted uncertainties on the data. This solution, however, also did not agree with our work nor \citet{triaud2020}. We repeated the analysis by imposing a prior on the orbital period and secondary eclipse time in Table~\ref{tab:variables} from our photometric solution. In this case the fit got even worse; the RV ``jitter'' term required to reconcile the two datasets is \qty{4}{\kilo\metre\per\second}.
We experimented with a range of non-Keplerian effects such as a linear and quadratic trend, changes in inclination ($\mathrm{d}i/\mathrm{d}t$), orbital period ($\mathrm{d}P/\mathrm{d}t$), and apsidal precession ($\mathrm{d}\omega/\mathrm{d}t$), none of which resulted in a satisfactory solution.
% added per referee commments:
Some potential causes for the high jitter value may be incorrect assumptions in our correction of the published RVs, or systematic noise sources not accounted for in the RV computation, such as wavelength calibration drifts over long baselines \citep[e.g.][]{sacco2014}, or the sensitivity of the derived RVs to the choice and treatment of spectral lines used in the analysis.
%%%%%%
A deeper investigation into whether the RV data can be explained by any other Keplerian signal or secular effect is outside the scope of this Letter. We conclude that the uncertainties on the RV data presented in \citet{baycroft2025} are underestimated by \qtyrange[range-units=single, range-phrase=\text{--}]{0.6}{4}{\kilo\metre\per\second} in its current form, and therefore the results derived from them cannot be compared to the photometric data in a meaningful way.
%% added based on discussions with baycroft et al.
Our conclusions are based on the assumption that our correction to the radial velocities is accurate, but should be confirmed by recomputing the RVs with consistent timestamps and with treatment of other sources of uncertainty.
%%% 

\subsection{Eclipse depth difference between SPECULOOS and TESS}
\label{section:depthdifference}

The phase folded eclipse in TESS is deeper than the SPECULOOS eclipse by about 4 percentage points at a significance of $2.6\sigma$, as shown in Figure~\ref{fig:mcmc}. Here we discuss a few possibilities for this apparent difference.

\textit{Contamination in TESS:} The eclipses in Sector 11 and latter half of Sector 65 occur near the edge of the light curves which coincides with the end of a TESS orbit. These regions are frequently associated with increased noise due to stray light from the Earth and the Moon. Light curves provided by \software{tglc} are corrected for the background and for contamination from nearby stars, however the correction may suffer as the background becomes brighter \citep{han2023}. This could result in an over-correction of the contamination flux, which could in principle lead to deeper eclipses. Due to the low S/N of individual eclipses we are not able to study their variation across sectors, but variation in flux contamination is thought be behind a range of depth discrepancies for both transiting exoplanets \citep{bryant2020a,han2025} and eclipsing binaries \citep{bryant2023} observed with TESS.

\textit{Contamination in SPECULOOS:} 
% The SPECULOOS light curve was derived from aperture photometry.
It is not currently known if the original SPECULOOS light curve was corrected for the flux from 2M1510\,C or whether it was necessary given the final aperture size. Given an atmospheric seeing of \qty{1}{\arcsec} and SPECULOOS' pixel scale of \qty{0.35}{\arcsec\per\pixel}, the FWHM of the PSF is about 2.9 pixels. Using a conservative $3\times\mathrm{FWHM}$ aperture radius gives a \qty{3}{\arcsec} aperture, makes it unlikely that 2M1510\,C at a separation of \qty{6.8}{\arcsec} contributes any significant flux into the aperture. On the off-chance that the real aperture was larger such that all of the light from 2M1510\,C fell into the aperture of 2M1510\,AB we can estimate what the true eclipse depth would be. 
We assume that the Gaia RP band is similar to the SPECULOOS $I+z'$, and estimate that given the Gaia RP magnitudes for the binary and tertiary ($G_\mathrm{rp} = 15.90$ and $17.3$) the flux ratio is $f_3/(f_1 + f_2) = 0.28$. The true eclipse depth of the binary would be roughly \qty{5.4}{\percent} instead of the currently observed \qty{4.2}{\percent}. This value does reconcile the two depths to within ${\sim}1\sigma$, although it is an upper limit of the effect.

\textit{Filter response differences:} We investigated whether the difference in the response functions between $I+z'$ and the TESS filter could account for the depth discrepancy, as the two brown dwarfs have slightly different effective temperatures. For this we calculated the total transmission for the SPECULOOS observations as a product of the $I+z'$ filter, the quantum efficiency curve of the detector, and atmospheric extinction using the ESO Sky Model Calculator\footnote{\url{https://www.eso.org/observing/etc/bin/gen/form?INS.MODE=swspectr+INS.NAME\%3DSKYCALC}}. We retrieved BT Settl atmospheric models of the two brown dwarfs assuming effective temperatures of \qty{2600}{\kelvin} and \qty{2700}{\kelvin}, with a $\log{g}=4.5$ and solar metallicity. We calculate the expected flux within the TESS filter and the SPECULOOS total throughput and find that the flux ratio of the two brown dwarfs in each filter differs only by \qty{1}{\percent} in relative terms, and can therefore not explain the difference in depth.

% \textit{Binary evolution:} While a change in orbital inclination could also explain the difference between the SPECULOOS and TESS data, we find this to be unlikely. The Kozai-Lidov effect can torque the inner orbit thereby changing the orbital plane of the binary. However in the case of 2M1510 the Kozai timescale is too long for this to be an observable effect \citep{triaud2020}. Moreover, the TESS data do not appear to show any significant differences in eclipse depth (nor duration) over the 6 years the system has been observed.

\section{Conclusion}
\label{sec:conclusion}
We have studied TESS full-frame image (FFI) photometry spanning 7 years of the double brown dwarf eclipsing binary 2M1510\,AB and detected a significant periodic signal using a box-least-squares (BLS) search. We find a new value
for the period $P = \qty{20.897782 \pm 0.000036}{\day}$ which is in agreement with the period from \citet{triaud2020}, whilst reducing its uncertainty from $\qty{8.5}{\minute}$ to $\qty{3}{\second}$. 
% This result is based on the significant peaks found in the two BLS runs at a period of $P = \qty{20.89776}{\day}$ with SDE values of 5.7 for SPECULOOS and TESS' lightcurves combined and 11.5 for TESS' lightcurves alone.
The TESS data favours a deeper eclipse (8\%) than SPECULOOS (4.2\%), which we cannot explain by filter differences nor orbital evolution due to the tertiary 2M1510\,C. However, we cannot rule out that the discrepancy arises from imperfect background correction of the original SPECULOOS or TESS images. Further, this only affects the shape of the lightcurve and not the position, and therefore does not affect the result for the period. Current data available at the time of writing does not indicate the presence of a circumbinary planet. Our work is crucial for scheduling follow-up observations of this system to allow for further study of the eclipse timing variations or secular evolution of the orbit, for example with JWST. 

\section*{Acknowledgements}
We thank the referee, Keivan Stassun, for his careful reading of the manuscript and for the constructive comments and suggestions which have helped improve the clarity and quality of the paper.
SM is grateful to the Department of Physics at the University of Warwick for funding through the Undergraduate Research Support Scheme (URSS). VK acknowledges funding from the Royal Society through a Newton International Fellowship with grant number NIF{\textbackslash}R1{\textbackslash}232229, and is grateful to Dan Bayliss, Ed Bryant, Tom Killestein, Tom Baycroft, Lalitha Sairam and Amaury Triaud for insightful discussions.

%%%%%%%%%%%%%%%%%%%%%%%%%%%%%%%%%%%%%%%%%%%%%%%%%%
\section*{Data Availability}

The TESS FFIs presented in this Letter can be accessed directly from the online Mikulski Archive for Space Telescope (MAST) portal\footnote{\url{https://mast.stsci.edu/portal/Mashup/Clients/Mast/Portal.html}}, while the processed light curves for 2M1510 are more easily accessed using the publicly available \software{tglc} software.\footnote{\url{https://github.com/TeHanHunter/TESS_Gaia_Light_Curve}} The SPECULOOS photometry was first published in \citet{triaud2020} and is available upon request.



\bibliographystyle{mnras}
\bibliography{example} % if your bibtex file is called example.bib

%\begin{thebibliography}{99}
%\end{thebibliography}

%%%%%%%%%%%%%%%%%%%%%%%%%%%%%%%%%%%%%%%%%%%%%%%%%%

%%%%%%%%%%%%%%%%% APPENDICES %%%%%%%%%%%%%%%%%%%%%

\appendix

\section{Predicted eclipse timings}

\begin{table}
    \sisetup{separate-uncertainty=true}
    \renewcommand{\arraystretch}{1.2}
    \small
    % \sisetup{round-mode=places}
    \centering
    \label{table:rotation_analysis}
    \caption{Predicted mid-point of secondary eclipses until the end of 2026. Uncertainties on each prediction range from 9-10 minutes. $^a$Since BJD$_\mathrm{UTC}$ 2457961.53255.}
    \begin{tabular}{@{\extracolsep{\fill}}
    lcc}
        \toprule
        \toprule
        Epoch$^a$ & Datetime & BJD \\
        & (UTC) & (UTC) \\
        \midrule
        144 & 22 Oct 2025 07:30:56 &  2460970.81316 \\ 
        145 & 12 Nov 2025 05:03:45 &  2460991.71094 \\
        146 & 03 Dec 2025 02:36:33 &  2461012.60872 \\
        147 & 24 Dec 2025 00:09:21 &  2461033.50650 \\
        148 & 13 Jan 2026 21:42:10 &  2461054.40429 \\
        149 & 03 Feb 2026 19:14:58 &  2461075.30207 \\
        150 & 24 Feb 2026 16:47:47 &  2461096.19985 \\
        151 & 17 Mar 2026 14:20:35 &  2461117.09763 \\
        152 & 07 Apr 2026 11:53:23 &  2461137.99541 \\
        153 & 28 Apr 2026 09:26:12 &  2461158.89320 \\
        154 & 19 May 2026 06:59:00 &  2461179.79098 \\
        155 & 09 Jun 2026 04:31:48 &  2461200.68876 \\
        156 & 30 Jun 2026 02:04:37 &  2461221.58654 \\
        157 & 20 Jul 2026 23:37:25 &  2461242.48432 \\
        158 & 10 Aug 2026 21:10:13 &  2461263.38211 \\
        159 & 31 Aug 2026 18:43:02 &  2461284.27989 \\
        160 & 21 Sep 2026 16:15:50 &  2461305.17767 \\
        161 & 12 Oct 2026 13:48:39 &  2461326.07545 \\
        162 & 02 Nov 2026 11:21:27 &  2461346.97323 \\
        163 & 23 Nov 2026 08:54:15 &  2461367.87102 \\
        164 & 14 Dec 2026 06:27:04 &  2461388.76880 \\
        \bottomrule
    \end{tabular}
\end{table}

\begin{table*} 
	\centering
	\caption{Orbital parameters and the physical characteristics used for the emcee fits, as well as their priors}
	\label{tab:jitters}
	\begin{tabular}{lccc}
		\toprule
        \toprule
		Nuisance parameters & Priors & TESS and SPECULOOS \\
		\midrule
        \textbf{Limb darkening Parameters $(u_1, u_2)$} \\
        SPECULOOS & 
        Fixed&
        $[0.26, 0.44]$
        \\

        TESS &
        Fixed&
        $[0.36, 0.45]$
        \\

        \textbf{log(f)}\\
        SPECULOOS & 
        $\mathcal{U}(-20.0, 0.0)$ & 
        $ -6.19 \pm 0.13 $  &
        \\

        Sector 11 & 
        $\mathcal{U}(-20.0, 0.0)$ & 
        ${<}{-1.2}$
        \\

        Sector 38 &
        $\mathcal{U}(-20.0, 0.0)$ &
        ${<}{-1.2}$
        \\

        Sector 65 (1st Transit) &
        $\mathcal{U}(-20.0, 0.0)$ &
        ${<}{-1.2}$
        \\

        Sector 65 (2nd Transit) &
        $\mathcal{U}(-20.0, 0.0)$ &
        ${<}{-1.2}$
        \\

        \textbf{Mean offset} \\
        SPECULOOS &
        $\mathcal{U}(0.8, 1.2)$ &
        $ 1.00004\pm0.00030 $
        \\

        Sector 11 &
        $\mathcal{U}(0.8, 1.2)$ &
        $0.9969\pm0.0071$
        \\

        Sector 38 &
        $\mathcal{U}(0.8, 1.2)$ &
        $1.0074\pm0.0043$
        \\

        Sector 65 (1st Transit) &
        $\mathcal{U}(0.8, 1.2)$ &
        $1.0008\pm0.0033$
        \\
        
        Sector 65 (2nd Transit) &
        $\mathcal{U}(0.8, 1.2)$ &
        $0.9998\pm0.0035$
        \\

        \textbf{Surface Brigthness}\\
        % SPECULOOS &
        % Fixed &
        % 0.827 
        % \\

        TESS &
        $ \mathcal{U}(0.0, 20.0) $ &
        $ 6.2_{-3.5}^{+7.3} $ 
        \\
        
        \bottomrule
		\hline
	\end{tabular}
\end{table*}

%\begin{table*} 
%	\centering
%	\caption{Priors for the two emcee runs. Orbital period, orbital inclination, sum of radii and secondary transit time were all varied, whilst radius ratio, eccentricity, argument of periastron, surface brightness ratio and mass ratio were all varied.}
%	\label{tab:priors}
%	\begin{tabular}{lcc}
%		\toprule
%       \toprule
%		Variables and Units & TESS Priors & TESS + SPECULOOS Priors \\
%		\midrule
%		Orbital Period ($\mathrm{days}$) &
%        $\mathcal{U}(20.87,20.93)$ &
%        $\mathcal{U}(20.87,20.93)$ &
%        \\
%        
%		$T_\mathrm{sec} (\mathrm{BJD - 2457000}$) & 
%        $\mathcal{U}(961.531,961.54)$ &
%        $\mathcal{G}(961.53518,0.00064)$ &
%        \\
%        
%	    Inclination $(^\circ)$ & 
%        $\mathcal{U}(88,90)$ &
%        $\mathcal{U}(60,90)$ &
%       \\
%       
%        Sum of Radii $(R_1 + R_2) / a~(\mathrm{R_\odot})$ &
%        $\mathcal{U}(0.178,0.583)$ &
%        $\mathcal{U}(0.0,1.0)$&
%         \\ 

%        Radius Ratio $k=(R_2/R_1)$ &
%        Fixed &
%        Fixed &
%       \\
        
%        Eccentricity &
%        Fixed &
%        Fixed & 
%        \\

%        Argument of Periastron $(^\circ$) &
%        Fixed &
%        Fixed &
%        \\
        
%        Surface Brightness Ratio $J_2/J_1 = k^2 f_2/f_1$ &
%        Fixed &
%        Fixed &
%        \\
        
%        Mass Ratio $M_1/M_2$ &
%        Fixed &
%        Fixed &
%        \\
        
%        \bottomrule
%		\hline
%	\end{tabular}
%\end{table*}

%%%%%%%%%%%%%%%%%%%%%%%%%%%%%%%%%%%%%%%%%%%%%%%%%%

% Don't change these lines
\bsp	% typesetting comment
\label{lastpage}
\end{document}